\input harvmac
\newcount\figno
\figno=0
\def\fig#1#2#3{
\par\begingroup\parindent=0pt\leftskip=1cm\rightskip=1cm
\parindent=0pt
\baselineskip=11pt
\global\advance\figno by 1
\midinsert
\epsfxsize=#3
\centerline{\epsfbox{#2}}
\vskip 12pt
{\bf Fig. \the\figno:} #1\par
\endinsert\endgroup\par
}
\def\figlabel#1{\xdef#1{\the\figno}}
\def\encadremath#1{\vbox{\hrule\hbox{\vrule\kern8pt\vbox{\kern8pt
\hbox{$\displaystyle #1$}\kern8pt}
\kern8pt\vrule}\hrule}}

\overfullrule=0pt

\Title{TIFR-TH/97-56, MIT-CTP-2694, MRI-PHY/P971030, hep-th/9711003}
{
 \vbox{
  \centerline{Boosts, Schwarzschild Black Holes and}
  \vskip 0.1 truein
  \centerline{Absorption cross-sections in M theory}
 }
}
\smallskip
\centerline{Sumit R. Das$^{(a)}$ \foot{E-mail: das@theory.tifr.res.in}, 
Samir D. Mathur$^{(b)}$ \foot{E-mail: me@ctpdown.mit.edu}, 
S. Kalyana Rama $^{(c)}$ \foot{E-mail: krama@mri.ernet.in} and 
P. Ramadevi$^{(a)}$ \foot{E-mail: rama@theory.tifr.res.in}}
\bigskip
\centerline{$^{(a)}${\it Tata Institute of Fundamental Research, 
Homi Bhabha Road, Bombay 400 005, INDIA}}
\centerline{$^{(b)}${\it Center for Theoretical Physics, Massachussetts 
Institute of Technology, Cambridge, MA 02139, USA}}
\centerline{$^{(c)}${\it Mehta Research Institute of Mathematics
and Mathematical Physics, Allahabad 221506, INDIA}}
\medskip

\medskip

\noindent $D$ dimensional neutral black strings wrapped on a circle are 
related to $(D-1)$ dimensional charged black holes by boosts. We show
that the boost has to be performed in the covering space and the
boosted coordinate has to be compactified on a circle with a Lorentz
contracted radius. Using this fact we show that the transition between
Schwarzschild black holes to black p-branes observed recently in M theory is
the well-known black hole- black string transition viewed in a boosted
frame. In a similar way the correspondence point where an excited
string state goes over to a neutral black hole is mapped exactly to
the correspondence point for black p-branes. In terms of the $p$ brane
quantities the equation of state for an excited string state becomes
identical to that of a $3+1$ dimensional massless gas for all $p$.
Finally, we show how boosts can be used to relate Hawking radiation
rates. Using the known microscopic derivation of absorption by
extremal 3-branes and near-extremal 
5D holes with three large charges we provide a
microscopic derivation of absorption of 0-branes by seven and five
dimensional Schwarzschild black holes in a certain regime.

\Date{October, 1997}

\newsec{Introduction}

Over the past year there has been a growing feeling that it is useful
to examine M theory in a boosted frame, to be better able to formulate
its structure and consequences.  In a recent set of papers this notion
was applied to attempt an understanding of Schwarzschild black holes
in various dimensions
\ref\bfks{T. Banks, W. Fischler, I. Klebanov and L. Susskind,
hep-th/9709091} \ref\ks{I. Klebanov and L. Susskind, hep-th/9709108.}
\ref\haylo{E. Halyo, hep-th/9709225.}.

The idea is to use the fact that neutral black holes become RR-charged
holes in a frame boosted along $x^{11}$. On the other hand, it has
been conjectured in \ref\susskind2{L. Susskind,
hep-th/9704080.} and shown in \ref\senseiberg{A. Sen, hep-th/9709220;
N. Seiberg, hep-th/9710009.}
that the hamiltonian of discrete light cone
quantization of M-theory is the finite $N$ version of M(atrix) theory
\ref\bfss{T. Banks,
W. Fischler, S. Shenker and L. Susskind, Phys. Rev. D55 (1997) 5112,
hep-th/9610043}. Thus DLCQ M(atrix) theory should give a
microscopic understanding of the thermodynamics of neutral black
holes in various dimensions. In \bfks\ this is done for $M$ theory
compactified on $T^3$ for which the DLCQ M(atrix) theory is the
well understood $N=4$ super-Yang-Mills theory in $3+1$ dimensions.
In \ks\ compactification on other $T^p$ have been considered
and the black hole results have been shown to be consistent with
known strong coupling results in M(atrix) theory for $p = 4$.

It is important to understand the precise benefits of
using the boosted frame.  When do we get something new, and when do we
simply get something already known but seen in a Lorentz transformed
frame? In this paper we ask the following questions :

\item{[A]} Suppose spacetime has a spatial compact direction. 
If we take a Schwarzschild black hole, and boost it in this compact direction,
then from the viewpoint of the noncompact space it acquires
Kaluza-Klein charge. Can we therefore relate some properties
of charged black holes to properties of neutral black holes?

\item{[B]} Which of the arguments in \bfks\ and \ks\ depend
on properties of M(atrix) theory and which of them follow from
known facts in general relativity and the kinematics of boosts ?

\item{[C]} Can we derive some interesting results about black holes which
were not otherwise calculable, by using the idea of boosts?

\noindent We have found the following answers to the questions above.

\item{[A]} \itemitem{(i)}
While the answer to the first question appears to be obviously `yes', 
we need to be
careful about the fact that boosting in a compact direction is not a
Lorentz symmetry. 
We need to discuss what boost transformation is required to 
establish a precise connection between neutral and charged black holes.  We
start with a ``black string'' in $(d+1)$ dimensions along $x^{d+1} =
z$ with no momentum along it. $x^{d+1}$ lies on a circle of radius
$R$.  Some of the other transverse directions may also be compact,
lying on a $T^p$ with each side of length $L$. Then this is a neutral
black hole in $(d-p)$ dimensions. We then perform a boost in the {\it
covering space} by an angle $\alpha$. Finally we compactify the new
$z$ coordinate $z'$ on a circle of radius $R'$. 
As a result we get a solution which after standard Kaluza
Klein reduction becomes an electrically charged $(d-p)$ black hole.
If $d = 10$ and the theory is M-theory, the charge is a 0-brane
charge of the ten dimensional string theory. In that case one can further
perform T-dualities to get a black D$p$-brane solution \ref\horstrom1{G.
Horowitz and A. Strominger, Nucl. Phys. B360 (1991) 197.}. 

\itemitem{(ii)}
The new point which we observe is that the resulting charged
black hole has the same entropy as the original neutral black hole
when $R' = R/\cosh \alpha$, i.e. the new radius is taken to be
precisely the Lorentz contracted radius.  This identifies the
precise nature of the ``boost transformation'' relevant to us.

\item{[B]}
\itemitem{(i)} 
We show that a crucial ingredient of \bfks\ and \ks - viz. that the
entropies of the boosted $(11-p)$ dimensional Schwarzschild black hole is
equal to the entropy of a black $p$-brane - is precisely the boosted
version of the black hole-black string transition
\ref\horopol{G. Horowitz and J. Polchinski, Phys. Rev. D55 (1997)
6189, hep-th/9612146; hep-th/9707170.} when the black hole radius
$\rho_0$ becomes greater than the radius of the transverse compact
direction $R$.  As such this fact is largely independent of any
property of the 11 dimensional M theory or M(atrix) theory; this
is a result in any theory which contains General Relativity in
$(d+1)$ dimensions. In M theory the charge is a 0-brane charge
of the string theory \ref\townsend{P. Townsend, Phys. Lett. B350
(1995) 184, hep-th/9501068.}. We can then
T-dualize the charged hole with $p$
compact transverse dimensions to a D-$p$ brane and use the microscopic
understanding of the resulting black D-$p$ brane.

\item{[C]}

\itemitem{(i)} 
We consider the correspondence principle formulated in \horopol\ which
says that string states at weak coupling transit to black holes at a
critical coupling, where both string and black hole descriptions have
the same entropy. We show that the correspondence point for neutral
black holes may be related to the correspondence point for black p-brane
(and hence to a large class of black holes with a single
large charge) by ``boosts'' described above.

\itemitem{(ii)}
For Schwarzschild black holes the weak coupling regime is described by
highly excited states of a single string. We rewrite the corresponding
equation of state in terms of the boosted variables and find that
for all values of $p$ the equation of state is that of a gas in
$3 + 1$ dimensions. This equation of state then gives a uniform
description of the weak coupling regime of black $p$-branes. Its
implications are not very clear at the moment.

\itemitem{(iii)}
We show that the above boosts may be used to relate the absorption
cross-sections of particles of different charges. In
particular we check that this procedure correctly predicts the
emission rate of charged scalars from the $4+1$ dimensional
near extremal black hole with three large charges from the emission rate
of neutral particles. This correspondence between rates is
independent of the way it is calculated (i.e. semiclassical or
microscopic) and simply a result of the kinematics of boosts.

\itemitem{(iv)}
We show that the relation between absorption/emission rates may be
now used to provide a microscopic description of absorption by a seven
dimensional Schwarzschild black hole - at least in a certain limit.  This is
because by boosts and T-duality we can go from this black hole to a
black 3-brane. In the extremal limit of the latter it is known
that the absorption of massless neutral scalars may be reproduced
exactly by a microscopic calculation in the $3+1$ dimensional
Yang-Mills theory
\ref\klebathree{I. Klebanov, Nucl. Phys. B496 (1997) 231,
hep-th/9702076; S. Gubser, I. Klebanov and A. Tseytlin,
Nucl. Phys. B499 (1997) 217, hep-th/9703040.}. By the kinematics
of boosts this
is related to absorption of certain charged particles by the
seven dimensional Schwarzschild black hole, in a suitable
limit.

\itemitem{(v)} In a similar way we argue that one can have a 
microscopic understanding of absorption or emission of certain
particles in five dimensional Schwarzschild black holes in string
theory. We show how such black holes are related to the well known
five dimensional black hole with three large charges by a chain of
boosts and T-dualities. Since absorption/emission cross-sections
in the latter has a microscopic derivation
\ref\fivedrefs{For reviews and references to the original
literature see S. D. Mathur, hep-th/9609053; 
G. Horowitz, hep-th/9704072; J. Maldacena, hep-th/9705078; 
; S.R. Das, hep-th/9709206.} we have a microscopic derivation 
for certain absorption/emission process in the former. In this
case one can deal with near-extremal rather than exactly extremal
holes so that we can make a statement about Hawking radiation.

\newsec{Boosted Schwarzschild black holes as charged black holes}

In this section we identify the precise nature of boost transformations
which relate black strings in $(d-p+1)$ dimensions, or equivalently
Schwarzschild black holes in $(d-p)$ dimensions, to charged black
holes in $(d-p)$ dimensions.

Consider pure Einstein gravity in $d+1$ space-time dimensions with a
Planck length $l_{pl}$. We will consider compactifications of this theory
to $(d-p)$ dimensions with the compact space being a torus $T^p \times
S^1$. The torus $T^p$ has volume $L^p$ while the radius of the circle
$S^1$ is R. In this theory there is a Schwarzschild black string
solution
\eqn\one{ds_{d+1}^2 = -(1-({r_0 \over r})^n)dt^2 
+ {dr^2 \over (1-({r_0 \over r})^n)} + r^2 d\Omega_{n+1} + dz^2
+ \sum_{i=1}^p (dx^i)^2}
where
\eqn\two{n = d-3-p}
and we have labelled the direction along $S^1$ by $z$.
By standard Kaluza-Klein reduction this is a black hole in $(d-p)$
dimensions.
The mass and entropy of this black hole are given by
\eqn\three{\eqalign{& M = {(n+1)\Omega_{n+1} L^p R \over 8 l_{pl}^{d-1}}~
r_0^n \cr
& S_{bs} = {\Omega_{n+1} L^p \pi R \over 2 l_{pl}^{d-1}}~r_0^{n+1}}}
where $\Omega_m$ denotes the volume of a unit $m$-sphere and the
$(d+1)$ dimensional Planck length $l_{pl}$ is defined in terms
of the $(d+1)$ dimensional Newton constant $G_{d+1} = l_{pl}^{d-1}$

We now go to the covering space (i.e. a noncompact $z$) and boost
along the $z$ direction by a boost angle $\alpha$. Thus
the coordinates in the boosted frame $(z',t')$ are
\eqn\five{\eqalign{&z' = z\cosh\alpha + t\sinh\alpha\cr
&t' = t\cosh\alpha + z\sinh\alpha}}
The metric now becomes
\eqn\six{\eqalign{ds_{d+1}^2 = & -(1 -{r_0^n\over r^n} \cosh^2\alpha)
(dt')^2
+ (1 + {r_0^n\over r^n} \sinh^2\alpha)(dz')^2 \cr
& + {r_0^n\over r^n}\sinh(2\alpha)~dz'~dt' \cr
& + {dr^2 \over (1-({r_0 \over r})^n)} + r^2 d\Omega_{n+1}
+ \sum_{i=1}^p (dx^i)^2}}
Finally we compactify the boosted coordinate on a radius $R'$. 
The momentum $P$ in the $z$ direction is then quantized and given
in terms of an integer $N$ by
\eqn\sixz{P = {N \over R'} = M{({R' \over R})} \cosh \alpha
\sinh \alpha}
with $M$ given by \three.

By standard Kaluza-Klein procedure the solution represents a charged
black hole in $(d-p)$ dimensions
\ref\horne{J.H. Horne, G. Horowitz and A.R. Steif,
Phys. Rev. Lett. 68 (1992) 568.}. The Einstein metric of this black hole
is given by
\eqn\kone{\eqalign{ds^2_{d} = [f(r)]^{({1 \over d-2})}&
\{-[f(r)]^{-1}
(1- {r_0^n \over r^n})(dt')^2 \cr
& + {dr^2 \over (1-({r_0 \over r})^n)} + r^2 d\Omega_{n+1}
+ \sum_{i=1}^p (dx^i)^2 \} }}
where
\eqn\ktwo{f(r) = 1 + {r_0^n \sinh^2\alpha \over r^n}}
The dilaton $\phi$ and the gauge field $A_0$ are given by
\eqn\kthree{\eqalign{& e^{2\phi} = [f(r)]^{({d-1 \over d-4})}\cr
& A_0 (r) = {r_0^n \sinh \alpha \cosh \alpha \over r^n + r_0^n \sinh^2
\alpha}}}
Note that the procedure which we have used is not a symmetry of the
theory. If we boost along a compact direction $z$, the new 
coordinate $z'$ is not compact with any radius. 
Such a procedure has been implictly used in earlier work on
classical solutions
\ref\odd{For review and references see A. Sen,
hep-th/9210005; G. Horowitz, in {\it String Theory and Quantum
Gravity, '92}, Proceedings of Trieste Speing School, 1992, 
ed. J. Harvey et.al (World Scientific, 1993); J. Russo and A. Tseytlin,
hep-th/9611047}. 

The Bekenstein entropy of this solution may be read off from
\kone\
\eqn\kfour{S_{ch} = 
{\Omega_{n+1} L^p \pi R' \over 2 l_{pl}^{d-1}}~r_0^{n+1}
\cosh\alpha}
while the ADM mass is given by
\eqn\kfive{M_{ADM} = M{({R' \over R})}[1 + {n \over n+1}\sinh^2\alpha]}
with $M$ given by \three.

The entropy, which should have an interpretation in terms of the
total number of states, should not change under boosts.
>From \kfour\ and \three\ it is clear that $S_{bs} = S_{ch}$ if
$R'$ is the Lorentz contracted radius
\eqn\ksix{ R' = {R \over \cosh \alpha}}
Note, however, that when viewed as black holes in $(d-p)$ dimensions, the
area of the $(d-p-2)$ dimensional horizon is {\it not} boost
invariant. The entropy is invariant since the $(d-p)$ dimensional
Newton's constant defined by
\eqn\kfived{G_{d-p} = {G_{d+1} \over L^p R}}
increases due to the contraction of $R$ to exactly compensate for
the increase of the area of $(d-p-2)$ dimensional horizon.

The momentum in \sixz\ transforms correctly
as may be seen by using \ksix\ to obtain
\eqn\keight{P = M \sinh \alpha}
as expected. 
The ADM mass \kfive\ can be seen to follow as 
a simple consequence of the Lorentz transformation of an 
appropriate $(d + 1)$-dimensional energy momentum tensor 
$T_{\mu \nu}$ which generates the metric \one\ asymptotically. 
We briefly outline the construction 
\ref\admgibb{G. W. Gibbons, Nucl. Phys. {\bf B207} (1982) 207.}. 

Setting $g_{\mu \nu} \equiv \eta_{\mu \nu} + h_{\mu \nu}$, 
the appropriate energy momentum tensor $T_{\mu \nu}$ is given by 
\eqn\admone{
R^{(1)}_{\mu \nu} - {1 \over 2} \eta_{\mu \nu} R^{(1)}  
= 8 \pi G T_{\mu \nu}}
where ($\partial_\lambda \equiv {\partial \over \partial x^\lambda}$) 
\eqn\admtwo{
R^{(1)}_{\mu \nu} = {1 \over 2} 
\left( \partial_\mu \partial_\nu h^\lambda_\lambda 
+ \partial_\lambda \partial^\lambda h_{\mu \nu} 
- \partial_\mu \partial_\lambda h^\lambda_\nu 
- \partial_\nu \partial_\lambda h^\lambda_\mu \right) 
}
is the part of the Ricci tensor linear in $h_{\mu \nu}$ 
and the indices are to 
be raised or lowered using $\eta_{\mu \nu}$ 
\ref\wein{S. Weinberg, {\it Gravitation and Cosmology}, 
John Wiley and Sons, Inc. (1972).}. Choosing the 
harmonic gauge where 
\eqn\admthree{
\partial_\lambda h^\lambda_\mu = {1 \over 2} \partial_\mu h 
\; , \; \; \; \;    \; \; 
h \equiv \eta^{\mu \nu} h_{\mu \nu} \; , 
}
the energy momentum tensor $T_{\mu \nu}$ is given by 
\eqn\admfour{
\partial_\lambda \partial^\lambda (h_{\mu \nu} 
- {1 \over 2} \eta_{\mu \nu} h) = 16 \pi G T_{\mu \nu} \; . 
}
The metric \one\ written in isotropic coordinates becomes 
\eqn\admfive{
d s_{d + 1}^2 = - \left( {4 \rho^n - r_0^n \over 4 \rho^n + r_0^n} 
\right)^2 d t^2 + \left( 1 + {r_0^n \over 4 \rho^n} 
\right)^{{4 \over n}} (d \rho^2 + \rho^2 d \Omega_{n + 1}^2) 
+ d z^2 + \sum_{i = 1}^p (d x^i)^2 
}
where $\rho$ is related to $r$ by 
\eqn\admsix{
2 \rho^n = r^n - {r_0^n \over 2} + \sqrt{r^{2 n} - r_0^n r^n} \; .
}
Then to the leading order in ${r_0^n \over \rho^n}$ we get 
$(a, \; b = 1, \; 2, \; \cdots p \; ; \; \; \; 
i, \; j = p + 1, \; p + 2, \; \cdots p + n + 2$) 
\eqn\admseven{
h_{00} = {r_0^n \over \rho^n} \; , \; \; \; \; 
h_{ij} = {r_0^n \over n \rho^n} \delta_{ij} \; , \; \; \; \; 
h_{zz} = 0 \; , \; \; \; \; 
h_{ab} = 0 \; , 
}
satisfying the harmonic gauge condition \admthree.
Note that $\rho^2 = \sum_{p + 1}^{p + n + 2} (x^i)^2$. 

Defining the mass $M$ through the relation 
$T_{00} = M \delta^{(n + 2)} (x^i)$ gives the expression 
for $M$ as in \three. Furthermore, it follows that 
\eqn\admeight{
T_{ij} = 0 \; ; \; \; \; \; 
T_{zz} = - {1 \over n + 1} T_{00} \; ; \; \; \; \; 
T_{ab} = - {1 \over n + 1} T_{00} \delta_{ab} \; . 
}
Now boost along the $z$ direction by a boost angle $\alpha$. 
Note that the Newton's constant $G$ increases to 
$G' = G \cosh \alpha$ due to the Lorentz contraction of $R$ 
as can be seen from equations \ksix\ and \kfived. 
Therefore, it follows from \admone\ that the combination 
$G T_{\mu \nu}$ must transform like a tensor. We thus get for 
the $00$-component 
\eqn\admnine{
G' T'_{00} = (1 + {n \over n + 1} \sinh^2 \alpha) \; G T_{00} \; . 
}
Defining the mass $M'$ in the boosted frame through the relation 
$T'_{00} = M' \delta^{(n + 2)} (x^i)$ gives 
\eqn\admten{
M' = M_{ADM} 
}
where $M_{ADM}$ is given in \kfive. 

All the results discussed so far are results in $(d+1)$ dimensional
General Relativity and has nothing to do with string theory. If
our starting point is 11-dimensional supergravity the dimensionally
reduced theory is Type IIA string theory. In that case we can perform
T-duality transformations to convert the charged black hole solution
into a Dirichelt p-brane solution. The identification of the entropy,
energy and momentum of course continue to hold.

\newsec{The black hole-black string transition}

In this section we demonstrate the relationship between the
neutral black hole - string transition \horopol\ to the 
boosted Schwarzschild black hole -  black $p$-brane transition
observed in \bfks\ and \ks. 

The object of interest in \bfks\ and \ks\ is the $(d+1-p)$ dimensional
Schwarzschild black hole, still keeping $z$ to be compact. Consider
such an object in its rest frame with horizon radius $\rho_0$.  The
solution is an array with a spacing $2\pi R$ along the covering
space of the circle along $z$ 
\ref\myers{D. Korotkin and H. Nicolai, gr-qc 9403029,
 Nucl. Phys. B429 (1994) 229, R. Myers, Phys. Rev. D35 (1987) 455.}. 
The mass and entropy of the black hole are
\eqn\lone{\eqalign{& {\tilde M} \sim { L^p \over l_{pl}^{d-1}}~
\rho_0^{n+1} \cr
& S_{bh} \sim {L^p  \over l_{pl}^{d-1}}~\rho_0^{n+2}}}
Consider now the $(d+1-p)$ dimensional Schwarzschild hole which has 
the same mass as that of the Schwarzschild string. From \three\
this means
\eqn\ltwo{({r_0 \over \rho_0})^n \sim ({\rho_0 \over R})}
The ratio of the entropies is
\eqn\ltwo{({S_{bs} \over S_{bh}}) \sim 
({\rho_0 \over R})^{1\over n}}
Thus for $\rho_0 <  R$ upto a numerical factor of order one,
the black hole array is entropically
favored while for $\rho_0 > R$ the black string is entropically
favored \horopol.
The common entropy at the black hole- black string transition
point $\rho_0 \sim R$ is
\eqn\lthree{S' \sim (RM)}
This transition is related to the instability of a black string
\ref\laflamme{R. Gregory and R. Laflamme, 
Phys. Rev. Lett. 70 (1993) 2837}. 
Some other properties of this transition have been
studied in \ref\mathura{S. D. Mathur, hep-th/9706151.}.

Now boost the system in the sense described above. The black
string becomes a $(d-p)$ dimensional charged black hole.
There will be a similar transition from the
boosted Schwarzschild black hole in $(d+1-p)$ dimensions and a charged
black hole in $(d-p)$ dimensions. Since the entropies do not change
under boosts the transition takes place once again when $\rho_0 \sim R$.
However, expressed in terms of the radius $R'$ after the boost this
point is
\eqn\lfour{\rho_0 \sim R' \cosh \alpha}
In \bfks\ and \ks\ it was assumed that the longitudinal size of
the $(d+1-p)$ dimensional Schwarzschild black hole contracts due to the
boost and the transition happens when this contracted radius is
of the same order as the radius of the circle. This gives precisely
the relation \lfour. The common entropy at the transition is
\eqn\lfive{S' \sim (N \coth \alpha)}
When the boost parameter is large 
one has a near-extremal charged black
hole. In this case $\coth \alpha \sim 1$  and  $S' \sim N$.

It is in general not easy to find the precise radius of
compactification where the entropy of the black hole will equal the
entropy of the black string carrying the same energy. The black hole
in the compact space can be thought of as an array in the covering
space, and the effective energy of each hole in the array is
influenced by the gravity of the other members of the array; thus the
metric of a single hole is not directly applicable to the array.

But by what was said above, we can find this critical radius of compactification
 for black holes carrying momentum charge, if we know it for neutral holes. Thus
 start with the neutral hole of mass $M$ in $d+1$ spacetime dimensions. Let the
compact circle have radius $R_c$ which is the point where the hole is unstable
 to formation of a black string stretching in the compact direction.
On dimensional grounds
$$R_c=\mu [G_NM]^{1/(d-2)}$$
where $\mu$ is a dimensionless constant of order unity, and $G_N$ is the
 gravitational constant in the $d+1$ dimensional spacetime.

In the covering space we see an array of black holes at spacing
 $R_c$. Now boost along the direction of the array, with boost angle
 $\alpha$. We now have an array of holes, each with mass
 $M\cosh\alpha$ and with momentm charge $P=M\sinh\alpha$. The array is
 of course still at the point of instability, but the separation
 between holes now appears as $R_c/\cosh\alpha$.

Thus we conclude that the critical radius for a hole with mass $M$ and
 charge $P$ will be
$$R_c=\mu [G_N M]^{1/(d-2)}[1-{P^2\over M^2}]^{(d-1)\over 2(d-2)}$$

Once again the results of this section are results in $(d+1)$ dimensional
General Relativity or any theory which contains it.

\newsec{The correspondence principle and boosts}

In this section we will restrict ourselves to $11$-dimensional M-theory
and the ten dimensional string theory which follows from it and
examine the correspondence principle of \horopol\ in the light
of boosts in the $11$th direction.

The string coupling $g_s$ and the string length $l_s$ are related to
the $11$ dimensional Planck length $l_{pl}$ and the size of the 11th
dimension $R$ by the well known relations
\eqn\aseven{l_s = ({l_{pl}^3 \over R})^{1\over 2}~~~~~~~~~~~~~~~~~
g_s = ({R \over l_{pl}})^{3\over 2}}

In \horopol\ a correspondence principle was formulated which
relates a weak coupling description in terms of a string state
with a strong coupling description in terms of a classical solution.
For neutral black holes the weak coupling description is in terms
of a highly excited state of a single string with an entropy
\eqn\asevenz{S_1 \sim m l_s} 
while at strong coupling the black hole entropy
is 
\eqn\asevenx{S_2 \sim m r_0}
where $m$ and $r_0$ are the mass and the radius
of the black hole. Thus the correspondence point is
\eqn\mone{r_0 \sim l_s}
At this point the curvature at the horizon is of the string scale.

For black D- $P$ branes the stringy description is in terms of a
Yang-Mills gas in $p+1$ dimensions for large $\alpha$ and in terms of
a single string state for small $\alpha$. The black brane description
is given precisely by the T-dual of the metric \kone. The correspondence
point once again occurs when the string metric curvature reaches the
string scale. This happens when \horopol\
\eqn\mtwo{r_0 \sim {{\tilde l_s} \over (\cosh \alpha)^{1\over 2}}}
where ${\tilde l_s}$ is the relevant string length.

However we saw that the D $p$-brane can be obtained by boosting a
neutral black string along $x^{11}$, i.e. a $(10-p)$ dimensional
Schwarzschild black hole and then performing T dualities. 
Thus the relevant string length ${\tilde l_s}$ in \mtwo\ has
to be identified with the string length in the boosted frame.
Because of the
boost the radius of $x^{11}$ changes from $R$ to $R'$, given by
\ksix. As a result the ten dimensional string theory obtained after
the boost has a different string length $l_s'$ and string coupling
$g_s'$ which may be read off from \aseven\
\eqn\mthree{\eqalign{&l_s' = ({l_{pl}^3 \over R'})^{1/2} = 
l_s  (\cosh\alpha)^{1\over 2} \cr
& g_s' = ({R' \over l_{pl}})^{3\over 2} = 
g_s (\cosh \alpha)^{-{3\over 2}}}}
Indeed, identifying ${\tilde l_s}$ in \mtwo\ with $l_s'$ in \mthree, 
we see that the correspondence point of neutral black holes
written in terms of the string scale in the boosted frame is
precisely the correspondence point for D $p$-branes given by
\mtwo.

To write the relations in terms of the string coupling and the volume
relevant to the $p$-brane description we perform a T-duality along the 
$T^p$ directions. The new string
coupling $g$ and the volume of the brane becomes
\eqn\mthreea{\eqalign{& g = g_s'({l_s' \over L})^p \cr
&V = \Sigma^p = ({l_s'^2 \over L})^p}}

For near-extremal D $p$-branes the stringy description is a Yang-Mills
gas. The energy of the gas $\Delta E$ has to be identified with the
excess energy of the black hole above extremality, which, from \kfive\ is
given by
\eqn\mthreeb{ \Delta E = M ({R' \over R})}
The ratio of the entropy
of the $p$-brane $S_p$ to that of the YM gas $S_g$ is
\eqn\mfour{{S_p \over S_g} \sim [(gN)
({\Delta E \over N^2 V})^{4-n
\over 8-n}]^{4-n \over 2n}}
which can be easily derived from the formulae given in \horopol. 

It may be easily checked using \mthreea, \mthreeb  and \mthree\ that the
formulae for the energy and entropy in Section 2. are the standard
formuale for D $p$-branes. In terms of the string coupling the 
correspondence point for the brane-gas transition is at $g = g_c$ where
\eqn\mseven{g_cN = ({\Delta E \over N^2 V})^{4-n
\over n-8}}
However, \mfour\ shows that the gas phase has more entropy for $g < g_c$
only when $n < 4$ ($p > 3$). 
For $n = 4$ ($p = 3$) one has $S_g \sim S_p$ for all
couplings, while $n > 4$ ($p < 3$) one has $S_p > S_g$ at weak
couplings. This pathological behavior for $p < 3$ 
is probably related to the
fact that for small values of $p$ there are strong infra-red
divergences in the worldbrane theory which makes the gas picture
unrealiable even at weak coupling.

On the other hand for $(10-p)$ dimensional neutral black holes the
system would be in the string phase at weak couplings for all values
of $p$. Since the black $p$-branes and the neutral black holes are
related by a boost, it is of some interest to see what the equation of
state for a string state becomes when expressed in terms of the
boosted variables.

When the boost $\alpha$ is very large (so that the boosted black hole is
near-extremal), it follows from \keight, \ksix\ and \sixz\ that
\eqn\meight{\eqalign{&M \sim {N \over R} \sim {N \over g_s l_s} \cr
& \Delta E \sim M e^{-\alpha}}}
Using \meight, \mthree, \mthreea\ and \mthreeb,
the entropy of excited string state
of mass $M$ in rest frame \asevenz\ may be expressed as
\eqn\mnine{S_1 \sim M l_s \sim [N^2(\Delta E)^3 ({l_s'^{3-p} \Sigma^p
\over gN})]^{1/4}}
Remarkably this is precisely the equation of state for a massless
gas in 3 space dimensions with $N^2$ degrees of freedom and a
volume
\eqn\anineteen{V = {l_s'^{3-p} \Sigma^p \over gN}}
and the result is true for all $p$. In the above we have expressed
everything in terms of the product $(gN)$ since this is the effective
open string coupling in the $p$-brane.

It may be easily checked that this expression for the microscopic
entropy agrees with the p-brane entropy at the correspondence point
and of course for all values of $p$ the entropy $S_1$ is greater than
$S_p$ at weak couplings.

We have reached an intriguing conclusion : an excited massive string
state appears like a three dimensional gas in a boosted frame. In
other words if we take such a massive string state and add some
0-brane excitations to it by a boost we get a three dimensional gas.
The reason for this is not known to us.

It has been problematic to reproduce the correct absorption 
properties of black holes in the context of correspondence principle
\ref\emparan{R. Emparan,
Phys. Rev. D56 (1997) 3591, hep-th/9704204.}
\ref\das3{S.R. Das, hep-th/9705165.} - at least when the naive guess
about the degrees of freedom in the stringy phase (which gives the
entropy correctly) is used. For black holes with two large charges, the
degrees of freedom has been correctly identified in \mathura\ and then
the absorption properties follow correctly. For black holes with a single
large charge this is not known yet. The above equation of state
may be useful in this regard.

\newsec{Emission and absorption using boosts}

In this section we show how to use boost transformations to relate
emission or absorption cross-sections for different processes.

\subsec{General formalism}

First we describe the general strategy. Consider a theory in 
${\tilde D}$ space-time dimensions with $q$ compact directions, one of
which we will label $z$ and consider a black hole in $D = {\tilde D}
-q$ dimensions which carries a Kaluza-Klein charge coming from 
momentum along $z$. Such a black hole may be obtained by boosting
(in the sense used in this paper) a neutral hole along $z$ by some
boost parameter $\alpha$. The coordinate radius of the black hole is
$r_0$.

Consider emission of a particle which is massless in the full
${\tilde D}$ dimensions with some momentum ${\vec k}$ along the
$(D-1)$ transverse noncompact spatial directions and some
momentum $e$ along $z$, but no momentum along the other compact
directions. From the $D$ dimensional perspective this
is a charged particle of charge $e$. The energy of the particle
$\omega$ is then
\eqn\none{\omega^2 = k^2 + e^2} 
Let the absorption cross-section of the particle be denoted by $\sigma
= \sigma (\omega,e;r_0,\alpha)$.  If the temperature of the black hole
is $T = T(r_0,\alpha)$, and $A_0 = A_0 (r_0,\alpha)$ is the
electrostatic potential at the horizon $r = r_0$, then the distribution
function for such particles is given by
\eqn\ntwo{\rho(\omega,e;r_0,\alpha) = [e^{\omega - eA_0 \over T} 
- 1]^{-1}}
The rate of emission of particles with energy between
$(\omega, \omega + d\omega)$ and charge between $(e, e+de)$ is given by
\eqn\nthree{\Gamma (\omega,e;r_0,\alpha)
= \sigma~~ \rho~~ ({d^{D-1} k \over (2\pi)^{D-1}})~(R de)}

Now boost the system further with some angle $\beta$. The boost
parameter of the new black hole obtained this way is then 
$(\alpha + \beta)$. The particle of energy and charge $(\omega, e)$
now becomes a particle of energy and charge $(\omega',e')$, where
\eqn\neight{\eqalign{&\omega' = \omega \cosh \beta + e \sinh \beta \cr
& e' = e \cosh \beta + \omega \sinh \beta}}
Let us
denote the rate of emission of these particles in the boosted
frame $\Gamma' (\omega',e';r_0,\alpha + \beta)$. We want to obtain
$\Gamma'$ from a knowledge of $\Gamma$.

In fact, $\Gamma$ and $\Gamma'$ are related by time dilation
factor
\eqn\nfour{\Gamma (\omega,e;r_0,\alpha) =
\Gamma' (\omega',e';r_0,\alpha+\beta)~{\cosh(\alpha + \beta)
\over \cosh (\alpha)}}

To obtain $\Gamma'$ we thus express $\Gamma$ in terms of the primed
variables. We make a change of variables from 
$(|k|,e) \rightarrow (\omega,e)$ so that
\eqn\nfive{ |k| d|k| de = \omega d\omega de}
and write the phase space factor as follows
\eqn\nfivea{d^{D-1}k de = d\Omega_{D-2}~|k|^{D-3} \omega d\omega de
= d\Omega_{D-2}~|k|^{D-3} \omega d\omega' de'}
since $ d\omega de = d\omega' de'$ under boosts and the transverse
momentum magnitude $|k|$ is unchanged. Furthermore
the new radius of the $z$ direction after the boost, $R'$ is related 
to $R$ by
\eqn\nsix{R = R'~{\cosh(\alpha + \beta)
\over \cosh (\alpha)}} 
The potential at the horizon of the black hole before the boost is
\eqn\nseven{A_0 = \tanh \alpha}
whereas the temperature is
\eqn\nnine{ T = {T_0 \over \cosh \alpha}}
The potential and the temperature of the boosted black hole are then
\eqn\neleven{\eqalign{&A_0' = \tanh (\alpha + \beta) \cr
& T' = {T_0 \over \cosh (\alpha + \beta)}}}
It then follows from \neight\ and \nseven\ and \neleven\ that
\eqn\nten{(\omega - e A_0) =(\omega' - e' A_0')  
{\cosh(\alpha + \beta) \over \cosh (\alpha)}}
so that
\eqn\ntwelve{\rho(\omega,e;r_0,\alpha) 
= \rho (\omega',e';r_0,\alpha+\beta) \equiv \rho'}
Thus we get, using \nfive,
\eqn\nfourteen{\Gamma(\omega,e;r_0,\alpha)
= {\sigma(\omega,e;r_0,\alpha) \over [e^{\omega' - e' A_0'
\over T'}-1]}~{\omega \over \omega'}~{d\Omega_{D-2}~|k|^{D-3} \omega' d\omega'
 R' de'
\over (2\pi)^{D-1}}~{\cosh(\alpha + \beta) \over \cosh (\alpha)}}
Thus we get, using \nfour\ and \nfivea,
\eqn\nfourteena{\Gamma'(\omega',e';r_0,\alpha + \beta)
= \sigma~({\omega \over \omega'})~\rho'~~({d^{D-1}k' \over (2\pi)^{D-1}})
~(R' de')}
Comparing with \nthree\ we therefore have
\eqn\nfifteen{\sigma'(\omega',e';r_0,\alpha + \beta)
= \sigma (\omega,e;r_0,\alpha) ~{\omega \over \omega'}}
which gives the absorption cross-section $\sigma'$ for particles
of energy $\omega'$ and charge $e'$ by the boosted black hole.

\subsec{Application to five-dimensional black holes with three charges}

A good check of our formalism is offered by emission of scalars from
the near-extremal five dimensional black holes with three charges
$Q_1,Q_5,N$ where $N$ is a Kaluza Klein charge and $Q_1$ and $Q_5$ are
D1-brane and D5-brane charges. The $1$-brane is wrapped on a circle
of radius $R$ and the $5$-brane is wrapped on a $T^4$ times this
circle. The volume of $T^4$ is $V$.

The boost parameter is then associated with the charge $N$ which is
the quantized momentum along $x^5$.  The general answer for the low
energy absorption cross-section for charged particles
\ref\gubkleba{S. Gubser and I. Klebanov, Nucl. Phys. B482 (1996) 173,
hep-th/9608108.} is known to be
\eqn\nsixteen{\sigma (\omega,e;r_0,\alpha) = A_H (r_0, \alpha)
~({\omega - e A_0 \over \omega})}
Here $e$ is the charge of the particle which is the momentum along
$x^5$ and
$A_H$ is the area of the three dimensional horizon given by
\eqn\nseventeen{A_H(r_0,\alpha) = A_H(r_0,0)~\cosh \alpha}
Note that this horizon area is {\it not} invariant under boosts
along $x^5$. However the entropy is invariant under boosts. This
is because 
\eqn\nsevena{S = {A_H \over 4 G_5}}
where $G_5$ is the five-dimensional Newton constant which is
related to the ten dimensional Newton constant by
\eqn\neighta{ G_5 \sim {G_{10} \over V R}}
Thus the Lorentz contraction of the radius $R$ results in an increase
of $G_5$ which exactly compensates for the expansion of $A_H$ implied
by \nseventeen. In other words the area of the {\it four} dimensional
horizon (which is the product of the {\it three} dimensional horizon
with the circle along $x^5$) is indeed boost invariant.

Now boost further along $x^5$ with a parameter $\beta$.
It immediately follows, using \nfifteen. \nsixteen\ and \nten\ that
\eqn\nnineteen{\sigma' (\omega',e';r_0,\alpha + \beta) = A_H (r_0, \
\alpha + \beta)
~({\omega' - e' A_0' \over \omega'})}
which is exactly what one would expect.

In particular we can choose $e = 0$ and obtain charged emission
from neutral emission without doing a separate calculation.
Similarly if we know how to calculate only neutral emission from
a neutral black hole ($\alpha = 0$) we could have used this formalism
to calculate the emission of charged particles from a charged black hole.

The important point to realize in all this is that the relation
between cross-sections derived above does not depend on {\it how}
the cross-sections are calculated in the first place - semiclassical
or microscopic. This is entirely given by the boost properties.

\subsec{Application to seven dimensional Schwarzschild black holes}

The above formalism may be used to obtain microscopic derivation of
absorption cross-section by a seven dimensional Schwarzschild hole in
string theory.

Consider a seven dimensional neutral black hole, or a black string in
eight dimensions. The corresponding 11 dimensional metric is given by
\one\ with $n = 4$. This hole may Hawking radiate particles
which are massless in 11 dimensions but carry some momentum in the
11th direction, i.e. 0 branes of the corresponding string theory.
The classical absorption cross-section may be calculated in principle.
However, {\it a priori} there is no obvious microscopic derivation
of the absorption cross-section.

Now boost the system along $x^{11}$ by some angle $\alpha$. The
black string now acquires a momentum. In the string theory language
this has become a 0-brane black hole with three compact direction. The
particles which are absorbed now have a different 0-brane charge
determined by boost properties. After T-duality we have therefore
a 3-brane absorbing 3 branes. 

We may further choose the boost parameter such that the $x^{11}$
momentum of the absorbed particles vanishes, so that after T-duality
we have a 3-brane absorbing massless neutral particles. 

When such particles are minimally coupled scalars, and the 3-brane is
exactly extremal, the classical absorption cross-section can be
exactly reproduced by a microscopic calculation in the $3+1$
dimensional Yang-Mills theory living on the brane 
\klebathree. Therefore, at least in a certain limit, 
we can use the boost properties
of absorption cross-sections to have a microscopic derivation of
absorption properties of certain ``0-branes'' from the seven dimensional
neutral black hole. 

The limit is of course peculiar, since the extremal 3-brane can be
obtained from \one\ by infinite boosting ($\alpha \rightarrow
\infty$) and then taking the limit of $r_0 \rightarrow 0$ keeping
\eqn\oone{\lambda^4 = r_0^4 \sinh^2 \alpha}
fixed. 

To see the precise limits consider the S-wave equation
of motion of a minimally coupled scalar of energy $\omega$ and
charge $e$ in the background of the original neutral solution
\eqn\otwo{{1\over r^5}\partial_r(r^5 g(r) \partial_r \chi)
+ {\omega^2 - e^2 g(r)\over g(r)}\chi = 0}
where
\eqn\othree{g(r) = 1 - {r_0^4 \over r^4}}
This scalar field has an energy $\omega'$ and charge $e'$ after the
boost. If we choose $e' = 0$ one therefore has
\eqn\ofour{\omega = \omega' \cosh \alpha
~~~~~~~~~e = -\omega' \sinh \alpha}
and the equation \otwo\ becomes
\eqn\ofive{{1\over r^5}\partial_r(r^5 g(r) \partial_r \chi)
+ {\omega'^2 \over g(r)}[1 + {\lambda^4 \over r^4}]\chi = 0}
where $\lambda$ is defined in \oone.

For $(\omega' r_0) << 1$ this is precisely
the equation which is solved in \klebathree\ to obtain the classical
cross-section for absorption of neutral scalars by extremal 3-branes.
In terms of $\omega, e$ and $m$ this regime of parameters is
\eqn\ofivea{\omega r_0 << (1- {e^2 \over \omega^2})^{-1/2}}
This guarantees that the {\it classical} cross-section of our
original problem is known once the classical cross-section
for the three brane is known. The latter has been obtained as an
expansion in the product $\omega' \lambda$. Written in terms of
the original parameters this product is
\eqn\osix{\omega'\lambda = (\omega r_0)[({e \over \omega})^2
( 1 - ({e \over \omega})^2)]^{1\over 4}}
which means that $\omega' \lambda$ can be kept small and finite
with $e \sim \omega$. 

The fact that the cross-section for the 3-brane can be reproduced
by a microscopic calculation shows that the cross-section for the
neutral black hole has a microscopic derivation in the regime
of parameters given by \ofivea\ and $e \sim \omega$.

\subsec{Application to five dimensional Schwarzschild black holes}

The above microscopic derivation of absorption by seven dimensional
Schwarzschild black holes holds only in the limit where the 3-brane to which
it is related by boosts and T-duality is {\it exactly} extremal.  We
now outline a  much cleaner example where one can once
again obtain an exact derivation of absorption of certain 0-branes
by a five dimensional Schwarzschild black hole in ten dimensions -
or a black string in eleven dimensions with five of the transverse
directions compact. This is related to the well known five dimensional
black hole with a 1-brane charge $Q_1$, a five brane charge $Q_5$
and some momentum along the 1-brane direction by the following
series of steps

\item{(i)} Start with the 11-dimensional metric corresponding to
black string along $x^{11}$ and the directions $(x^1 \cdots x^5)$
compact.

\item{(ii)} Perform a boost along $x^{11}$ by a parameter $\gamma$.
The ten dimensional description of this is a $0$-brane with 
five compact directions with a charge $Q_5 \sim r_0^2 \sinh (2\gamma)$

\item{(iii)} Perform a T-duality along the directions $(x^1 \cdots x^4)$.
The $0$ brane now becomes a $4$-brane with the same charge $Q_5$

\item{(iv)} Consider the $11$-dimensional object whose ten dimensional
description is given by (iii). Perform a boost along $x^{11}$ with
some parameter $\alpha$. The resulting ten dimensional object is
a $4$-brane with some additional $0$-brane charge $Q_1 \sim
r_0^2 \sinh (2\alpha)$, and a compact transverse direction $x^5$.

\item{(v)} Now perform a T-duality along $x^5$. The $4$-brane 
becomes a $5$-brane with the same charge $Q_5$ and the $0$-brane
becomes a $1$-brane with the charge $Q_1$.

\item{(vi)} Finally perform a boost along the compact direction
$x^5$. One now has the well known system of $1$-branes, $5$-branes
and momentum. 

Following the above chain of boosts and dualities it is easy to check
that the metric obtained starting from the standard metric \one\ is
the metric of $1$-brane - $5$-brane system obtained in
\ref\formalstr{M. Cvetic and D. Youm, Nucl.Phys. B476 (1996) 118,
hep-th/9603100} and
\ref\hormalstrom{G. Horowitz, J. Maldacena and A. Strominger,
Phys.Lett. B383 (1996) 151, hep-th/9603109.}. 
The $11$-dimensional metric may be viewed as intersecting
M-branes \ref\intmbrane{For a review and references to the original
literature see J. Gauntlett, hep-th/9705011.}. The above steps have
been discussed earlier in \ref\tsyet{M. Cvetic and A. Tseytlin,
Nucl. Phys. B478 (1996) 181; A. Tseytlin, Phys. Lett. B 395 (1997) 24.}
\foot{We thank A. Tseytlin for pointing out these papers to us}.
Thus the absorption
of certain $0$-brane charged objects by the
five dimensional Schwarzschild black hole is related
to absorption of neutral objects by the five dimensional black hole
with three charges.

The low energy absorption cross-section for the 5D black hole with three
charges is, however, {\it exactly} reproduced by a microscopic 
calculation  not just in the extremal limit, but
in the near-extremal situation as well
\fivedrefs. In other words there is
a microscopic derivation of the rate of Hawking radiation for this
case. This implies that we have a microscopic derivation of
Hawking radiation of related particles from a five
dimensional Schwarzschild as well.

Note that the T-dualities and the boosts in the above chain do not
commute. This makes the identification of the particles emitted from
the Schwarzschild black hole which are related to neutral particles in the
5D black hole with three large charges rather involved. A more
detailed discussion of emission amplitudes will appear in a future
manuscript.
\vskip 1.0cm
\noindent{\bf{Note added}}

While this paper was being typed we saw \ref\horomart{G. Horowitz
and E. Martinec, hep-th/9710217} which has some overlap with
Section 2. of our work. Another related paper is \ref\miao{M. Li, hepth 9710226.}

\vskip 1.0cm
\noindent{\bf{Acknowledgements}} 

We would like to thank A. Dabholkar and
P. Joshi for discussions. S.K.R. would like to thank the Theoretical
Physics Group of Tata Institute of Fundamental Research for
hospitality.  S.D.M was partially supported by DOE cooperative
agreement No.  DE-FC02-94ER40818.

\listrefs
\end